\documentclass[11pt]{article}

\usepackage[T1]{fontenc}
\usepackage{lmodern}
\usepackage[margin=25mm]{geometry}
\usepackage{amsmath,amssymb}
\usepackage{graphicx}
\usepackage{booktabs}
\usepackage{tabularx} %
\usepackage{siunitx}
\usepackage[version=4]{mhchem}
\usepackage{xcolor}
\usepackage{listings} %
\usepackage[numbers,super,sort&compress]{natbib}
\usepackage{lineno}
\usepackage[hidelinks]{hyperref}
\usepackage{authblk}

\newif\ifarxiv
\arxivtrue

\makeatletter
\newcommand{\setsuppnumber}[2]{\ifarxiv\else\@namedef{r@#1}{{#2}{}{}{}{}}\fi}
\makeatother
\setsuppnumber{supp:note-prompts}{1}
\setsuppnumber{supp:note-reasoning}{2}
\setsuppnumber{supp:fig-size}{S1}
\setsuppnumber{supp:tab-models}{S1}
\setsuppnumber{supp:tab-descriptions}{S2}

\newcommand{\TD}{Typed Decision}
\newcommand{\SO}{Structured Outputs}
\newcommand{\RSO}{Reasoning+Structured Outputs}
\newcommand{\dose}{\ensuremath{\mathrm{e}^{-}/\text{\AA}^{2}}}

\title{Fast closed-loop STEM navigation \\
using vision-language models with typed decisions}
\author{Shuichiro Ozawa}
\author{Izumi Takahara}
\author{Fumihiko Uesugi}
\author{Teruyasu Mizoguchi\textsuperscript{*}}
\affil{Institute of Industrial Science, The University of Tokyo, 153-8505 Tokyo, Japan\protect\\[0.5ex]
\textsuperscript{*}Corresponding author. E-mail: \href{mailto:teru@iis.u-tokyo.ac.jp}{teru@iis.u-tokyo.ac.jp}}
\date{}

\begin{document}
\maketitle

\begin{abstract}
Closed-loop scanning transmission electron microscopy (STEM), in which each acquired image determines the next field of view and magnification, offers a route to autonomous atomic-scale materials analysis. 
Such a loop would automate navigation to defects and grain boundaries, a task that has long relied on operator judgment, and would enable systematic analysis of diverse local structures at scale. 
Realizing it hinges on the decision step, which must translate each acquired image into the next operation both quickly and correctly; yet which decision method can deliver both remains unclear. 
Here, we isolate this decision step.
Using simulated atomic-resolution STEM images of a substitutional W dopant and a grain boundary in monolayer \ce{MoS2}, where the true target position is known and every method starts from identical conditions, we compared four vision-language models (VLMs) and three decision methods. 
Our results demonstrate the effectiveness of \TD{}, which directly obtains the probabilities of candidate labels from a pretrained VLM, for closed-loop STEM navigation. 
\TD{} showed the shortest decision latency for all models (0.13--0.21~s) and generally achieved the highest navigation success rate among the three methods: it was the best for all models in W dopant navigation (66--91\%) and for the best-performing model, Cosmos-Reason2 32B, in grain boundary navigation (43\%). 
Furthermore, navigation to the different targets, the W dopant and the grain boundary, was possible by natural-language specification alone, without task-specific training of the VLM. 
These results indicate that \TD{} is an effective decision method for realizing fast, accurate, and flexible closed-loop STEM.
\end{abstract}

\section{Introduction}

\begin{figure}[tbp]
  \centering
  \includegraphics[width=\linewidth]{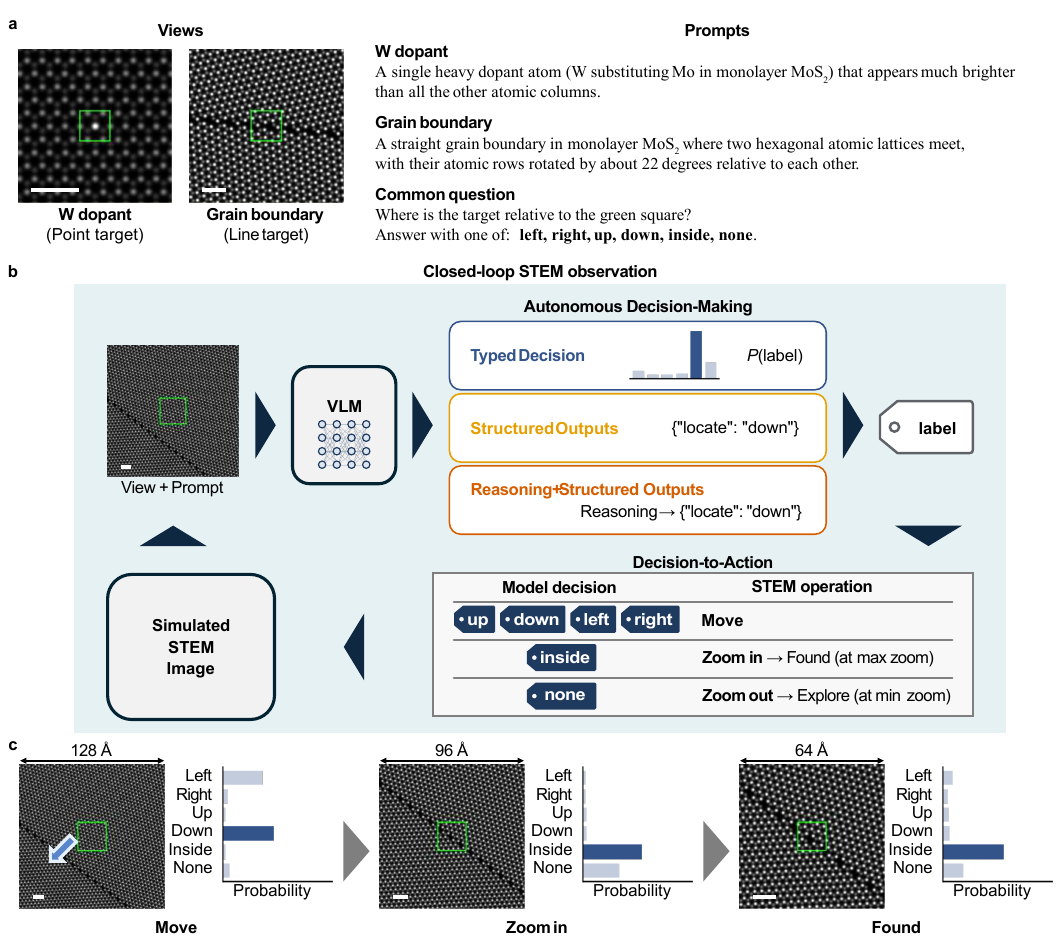}
  \caption{
  \textbf{Overview of closed-loop STEM navigation using typed decisions by a VLM.} 
  \textbf{a}, Views seen by the model in the W dopant and grain boundary tasks.
  The W atom ($Z = 74$) appears brighter than the surrounding Mo atoms ($Z = 42$).
  The Prompts are input to the model together with the images. 
  \textbf{b}, Overview of the closed loop based on typed decisions by the VLM. 
  \textbf{c}, Example of grain boundary navigation with \TD{} (movement based on the predicted label probabilities).
  The FOV width is shown above each image, the bar charts show the predicted label probabilities, and the arrow indicates the direction of FOV movement.
  Scale bars in the STEM images, 1~nm.
  }
  \label{fig:overview}
\end{figure}

Electron microscopy (EM) is widely used to analyze the microstructure and local properties of materials, and scanning transmission electron microscopy (STEM) plays an important role in atomic-scale imaging and chemical analysis\cite{Susi2022-oh,Ophus2023-jc}. 
In materials observation by EM, the position of the field of view (FOV) and magnification must be adjusted to navigate to specific atoms or defects and to perform local analysis. 
Conventionally, an expert operator checks the observed images on the display, interprets them, and sequentially adjusts the FOV position and measurement conditions, moving the specimen stage to the region of interest\cite{Kalinin2021-ko,Spurgeon2021-rx}.
Although routine procedures such as instrument alignment have been increasingly automated\cite{Spurgeon2021-rx}, decisions on specimen movement and lens adjustment (e.g., magnification, focus, and astigmatism), as well as the interpretation of the observed images, still depend largely on the expertise and experience of the operator\cite{Kalinin2023-ys,Jamali2026-bf}.

Meanwhile, self-driving laboratories, which run the entire process of materials research without human intervention, from synthesis, measurement, and analysis to the proposal of the next experiment, are being actively developed\cite{Burger2020-zb,MacLeod2020-zs,Tom2024-yh,Szymanski2023-al,Takahara2026-hy}. 
In such a process, the reliance on expert operators becomes a bottleneck, because the process is interrupted whenever the measurement stage requires their judgment and operation.
To remove this bottleneck, the microscope itself must decide the next observation from each acquired image and operate the instrument accordingly, so that measurement proceeds without an operator.
In this study, we refer to this mode of operation, in which acquiring a STEM image, determining the next FOV position and magnification on the basis of that image, and operating the instrument are repeated without an operator, as \emph{closed-loop STEM}.
Closed-loop STEM thus serves as an essential stage of self-driving laboratories, and realizing it requires observation planning that determines what to observe, sequential decision-making that rapidly selects the next operation from each acquired image, and a mechanism that converts these decisions into instrument operations.

To date, several machine-learning-based approaches to autonomous EM have been reported\cite{Creange2022-mh,Addiego2025-gr,Roccapriore2022-nm}.
Creange et al.\ developed workflows in which Bayesian optimization selects sampling points guided by features extracted with a pretrained convolutional neural network (CNN) and autoencoders, and evaluated them on pre-acquired images\cite{Creange2022-mh}.
Addiego et al.\ demonstrated autonomous control of the FOV position and magnification based on CNN-based object detection\cite{Addiego2025-gr}. 
Roccapriore et al.\ realized sequential learning of structure--property relationships and autonomous selection of electron energy-loss spectroscopy measurement positions using deep kernel learning\cite{Roccapriore2022-nm}.
In these systems, however, the operator must define what to look for in advance for each task.
The target is defined either by training a model on labeled images of it\cite{Addiego2025-gr} or by hand-designing a quantity for the system to maximize, such as a latent variable selected from image features or a scalar computed from each spectrum\cite{Creange2022-mh,Roccapriore2022-nm}.
The observation target therefore cannot be specified directly, for example in natural language.
Retargeting these systems to a different feature also requires new labeled training data or a newly designed quantity to be maximized.

More recently, artificial intelligence (AI) agents based on large language models (LLMs) and vision-language models (VLMs) have extended microscope automation beyond the recognition of individual images\cite{Ziatdinov2017-te, Maksov2019-ff, Yang2021-ts, Akers2021-xh}, allowing observation goals to be specified in natural language and experimental operations to be executed autonomously\cite{Yang2025-nj}.
For example, LLM agents have been used to link experimental instructions and plans to specific measurements and operations\cite{Wall2026-uu,Du2026-po}. 
Applications to image interpretation have also advanced, including the analysis of electron microscopy images with multiple agents and the evaluation of sequential observation decisions by CNNs and VLMs\cite{Chen2026-pr,Polat2026-bi}. 
These studies suggest that AI could handle tasks ranging from understanding observation goals to image analysis and experimental operation. 

However, in closed-loop STEM, the decision on the next operation is made at every step, often tens of times per target, while the specimen continues to accumulate electron dose and drift. 
A decision slower than image acquisition, which takes several seconds per atomic-resolution frame, therefore increases beam damage and lets the FOV drift away from the image on which the decision was based\cite{Egerton2004-zt, Komsa2012-ml}. 
Speed is thus required in addition to accuracy, yet how the choice of decision method trades one against the other has not been systematically examined.

VLM-based agents usually make each decision by generating text, often after an explicit reasoning trace, and this generation takes time because the output is produced token by token.
A faster alternative is found in the evaluation of VLMs on multiple-choice image understanding, where an answer is selected by scoring the likelihood of each candidate answer or its token scores\cite{Alayrac2022-qz,Wang2024-vi, Li2024-ki}.
Applied to microscopy, this scoring offers three properties suited to iterative decisions: a single forward pass replaces autoregressive generation; the output is always one of the predefined labels and can be mapped directly onto a microscope operation; and the label probabilities quantify the confidence of each decision. 
Recently, models that return probabilities over predefined labels without generating text have been introduced as typed decision models\cite{Almeida2026-sy,Tang2026-td}.
We refer to obtaining decisions in the same way from a pretrained VLM as \emph{\TD{}}.
\TD{} is thus expected to be fast, but how accurate it is in closed-loop navigation remains unclear, because label probabilities do not always agree with the answers that the same model generates as text\cite{Wang2024-vi}.

In this study, we propose closed-loop STEM navigation that incorporates a decision model, which selects one of a predefined set of labels from each acquired image, and converts the selected label into a microscope operation according to fixed rules.
Here, we use pretrained VLMs with \TD{} as the decision model, so that the observation target can be specified in natural language.
For comparison, we replace \TD{} with two decision methods based on text generation, the standard way in which VLM-based agents make decisions.
\SO{} has the model generate its answer as text in a constrained output format that allows only the same labels (Fig.~\ref{fig:overview}b).
\RSO{} further generates an explicit reasoning trace before the answer, which is widely expected to improve decision accuracy\cite{Wei2022-mz, Yao2022-vs, Zhang2023-xo, Zawalski2024-it}.
By comparing \TD{} with these methods, we examine whether \TD{} can make decisions much faster without losing accuracy relative to text generation, including generation with explicit reasoning.

We test the proposed closed-loop navigation on two representative defects in simulated monolayer \ce{MoS2}, a prototypical two-dimensional material\cite{Splendiani2010-ep,Manzeli2017-td}.
One is a point-like substitutional W dopant, and the other is a line-like grain boundary\cite{van-der-Zande2013-hi,Zhou2013-ub}.
For each method, we evaluate the decision latency, navigation success rate, and causes of failure.
We further examine the applicability and limitations of our approach through the effects of observation noise and changes in the target description.
This study presents a practical approach to fast and flexible closed-loop STEM navigation that connects decisions obtained directly from pretrained VLMs to microscope operations.

\section{Results}

\subsection{Closed-loop navigation setup} 

Using simulated HAADF-STEM images of monolayer \ce{MoS2}, we set up two navigation tasks: one targeting a W dopant and the other targeting a grain boundary (Fig.~\ref{fig:overview}a). 
For the W dopant task, we used a structure in which one Mo atom in \ce{MoS2} was replaced by a W atom.
Because W has a higher atomic number than Mo ($Z = 74$ vs.\ 42), the W atom appears in HAADF-STEM images as a bright spot with higher intensity than the surrounding Mo atoms\cite{Krivanek2010-hg, Pennycook1991-hm, Dumcenco2013-vq} (Fig.~\ref{fig:overview}a, left).
For the grain boundary task, in contrast, we used a structure in which two regions with crystal orientations differing by approximately $22^{\circ}$ are joined\cite{Najmaei2013-qj, Yazyev2010-hw}. 
In HAADF-STEM images, the grain boundary appears as a boundary at which the lattice orientation changes\cite{van-der-Zande2013-hi, Zhou2013-ub} (Fig.~\ref{fig:overview}a, right). 
For both the point-like W dopant and the line-like grain boundary, the goal was to bring the target inside a green square at the center of the FOV at the highest magnification by adjusting the FOV position and magnification.
Noise-free images simulated at exact focus ($\Delta f = 0$) were used here to compare the decision methods under controlled conditions; the effect of shot noise at finite electron doses is examined in Sec.~\ref{sec:variability} (Fig.~\ref{fig:robustness}a,b).

An overview of the closed-loop navigation is shown in Fig.~\ref{fig:overview}b. 
At each step, the VLM received two inputs: the current STEM image with a green square overlaid at the center of the FOV, and a text prompt.
The green square is a reference region placed on the image for judging the position of the target\cite{Shtedritski2023-wg}. 
The prompt described the W dopant as an atom brighter than its surroundings and the grain boundary as a boundary with a misorientation of approximately $22^{\circ}$, and instructed the VLM to judge where the target was located relative to the green square (Fig.~\ref{fig:overview}a, Prompts). 
Specifically, the VLM was asked to answer \texttt{inside} if the target was within the green square; one of \texttt{up}, \texttt{down}, \texttt{left}, or \texttt{right}, indicating the relative direction, if the target was visible outside the green square; and \texttt{none} if the target was not visible in the image. 
The full prompts are provided in Supplementary Note~\ref*{supp:note-prompts}.

The label output by the VLM was converted into a STEM operation according to predefined rules (Fig.~\ref{fig:overview}b, Decision-to-Action). 
When a direction label was output, the FOV position was moved in that direction; in \TD{}, the direction was weighted by the predicted probabilities of the four direction labels (Eq.~\ref{eq:move}).
For \texttt{inside}, the magnification was increased, and navigation was terminated when the target was judged to be \texttt{inside} at the highest magnification. 
For \texttt{none}, in contrast, the magnification was decreased, and navigation was terminated when \texttt{none} was returned at the lowest magnification.
The starting positions were chosen such that the target came into view at the lowest magnification (see Methods).
By repeating image acquisition, position judgment by the VLM, and STEM operation in this way, we performed closed-loop navigation toward the target. 

An example of grain boundary navigation is shown in Fig.~\ref{fig:overview}c. 
In the initial FOV, the grain boundary was located to the lower left of the green square.
Because the predicted probabilities were split between \texttt{down} and \texttt{left}, the FOV was moved diagonally toward the grain boundary rather than along a single axis (Eq.~\ref{eq:move}).
After the grain boundary entered the green square, the magnification was increased, and navigation was terminated when the grain boundary was again judged to be \texttt{inside} at the highest magnification.

We used four pretrained VLMs (Cosmos-Reason2, Qwen3.8, Gemma 4, and Nemotron 3 Nano Omni; Supplementary Table~\ref*{supp:tab-models}) without task-specific training. 
Because the weights of all four models are publicly available, we could read out the output probability of each label directly.
For these same models, we also compared the navigation success rate and decision latency of three decision methods (Fig.~\ref{fig:overview}b): \TD{}, which selects a label from the output probabilities of the labels; \SO{}, which generates a label in a constrained output format; and \RSO{}, which generates an explicit reasoning trace before the label.

For detailed analysis of the episode outcomes and navigation trajectories in grain boundary navigation, we used Cosmos-Reason2 32B, which performed best among the four VLMs.
This model was developed for temporal and spatial understanding and robot planning\cite{NVIDIA2026CosmosReason2}, and is therefore suited to the present task, which requires judging the relative position of the target within the FOV.
We further evaluated the robustness of \TD{} with Cosmos-Reason2 32B against observation noise (image-side variability; Fig.~\ref{fig:robustness}a,b) and changes in the target description (language-side variability; Fig.~\ref{fig:robustness}c).

\subsection{Comparison of decision methods} 
\label{sec:comparison}

\begin{figure}[tbp]
  \centering
  \includegraphics[width=\linewidth]{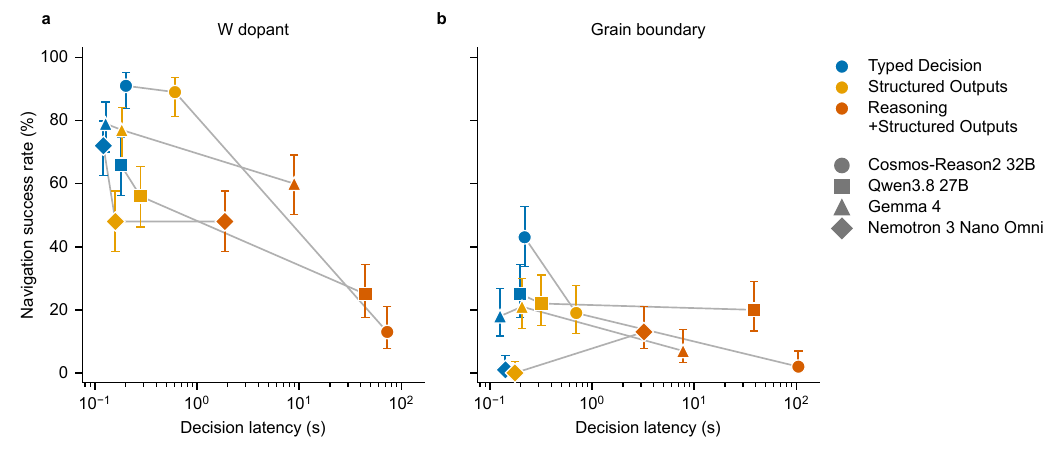}
  \caption{
  \textbf{Navigation success rate and decision latency of closed-loop navigation with different decision methods.}
  The four VLMs were compared using \TD{}, \SO{}, and \RSO{} for W dopant and grain boundary navigation. 
  Each condition consisted of 100 episodes with randomized starting positions and was evaluated without noise and with a step limit of 60 steps. 
  Error bars indicate 95\% Wilson confidence intervals, and gray lines connect the three decision methods for the same VLM. 
  \textbf{a}, W dopant navigation. 
  \textbf{b}, Grain boundary navigation.
  }
  \label{fig:methods}
\end{figure}

To clarify how the decision method affects closed-loop navigation, we evaluated two metrics.
The navigation success rate is the fraction of episodes in which the target was correctly judged to be inside the green square at the highest magnification, and the decision latency is the time required for a single decision.
As shown in Fig.~\ref{fig:methods}, \TD{} had the shortest decision latency for all models, at 0.13--0.21~s per decision.
Depending on the model, \SO{} and \RSO{} took 1.3--3.1 and 21--397 times longer per decision, respectively.

In contrast, the success rate of closed-loop STEM navigation (i.e., the success rate of the task of bringing the W atom or the grain boundary inside the green square) varied with the task, model, and decision method. 
In the W dopant task, \TD{} achieved the highest navigation success rate for all models (66--91\%). 
In the grain boundary task, the success rates were lower than those in the W dopant task for all models and decision methods. 
The mean success rate was 60\% for the W dopant task and 16\% for the grain boundary task, approximately one-quarter of that for the W dopant task. 
This ratio was similar across decision methods.

In the grain boundary task, the most effective decision method depended on the model.
With Cosmos-Reason2 32B, \TD{} achieved the highest success rate (43\%), whereas \RSO{} showed a low success rate (2\%). 
With Nemotron 3 Nano Omni, in contrast, \TD{} achieved 1\% and \RSO{} achieved the highest success rate (13\%).
Within Cosmos-Reason2, reducing the model size from 32B to 2B shortened the decision latency of \TD{} to about one-third (from 0.21 to 0.07~s) but lowered the navigation success rate from 91\% to 9\% for the W dopant and from 43\% to 1\% for the grain boundary (Supplementary Fig.~\ref*{supp:fig-size}).
Taken together, \TD{} was consistently the fastest, whereas the most successful decision method depended on the task and model.
Based on these results, the next section examines the types of failure that occurred with each decision method by analyzing the episode outcomes and individual failure cases.

\subsection{Failure analysis} 
\label{sec:failure}

\begin{figure}[tbp]
  \centering
  \includegraphics[width=\linewidth]{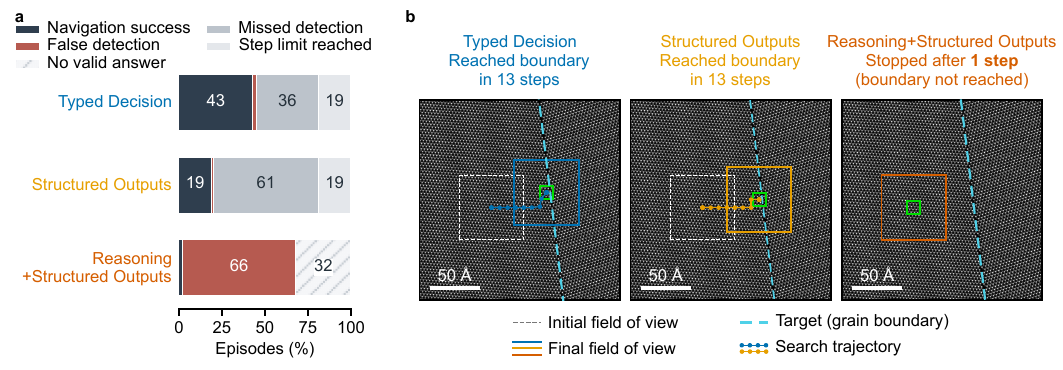}
  \caption{
  \textbf{Episode outcomes and failure cases of closed-loop navigation.} 
  \textbf{a}, Breakdown of the episode outcomes of the three decision methods in grain boundary navigation with Cosmos-Reason2 32B. 
  Each condition was evaluated with the same 100 episodes as in Fig.~\ref{fig:methods}. 
  Episodes were classified by how they ended: navigation success (\texttt{inside} at the highest magnification with the target inside the green square), false detection (\texttt{inside} at the highest magnification with the target outside the green square), missed detection (\texttt{none} at the lowest magnification with the target within the FOV), step limit reached (no termination within 60 steps), and no valid answer (no answer in the specified format).
  \textbf{b}, Navigation trajectories of grain boundary navigation with Cosmos-Reason2 32B starting from the same initial position. 
  \TD{} and \SO{} both succeeded in 13 steps, whereas \RSO{} incorrectly judged the target to be inside the green square at the first decision and terminated navigation (False detection). 
  The white dashed line indicates the initial FOV, the colored solid lines indicate the final FOVs, the green squares indicate the final position of the green square, and the light blue dashed line indicates the target (grain boundary). 
  The dotted lines with markers indicate the search trajectories (the FOV center at each step).
  For \RSO{}, the final FOV coincides with the initial FOV because navigation was terminated at the first decision.
  The white dashed line, the colored solid lines, the search trajectories, and the light blue dashed line are not part of the STEM images used in the analysis; they are drawn in this figure only to highlight each region.
  }
  \label{fig:failure}
\end{figure}

To examine in detail how the navigation success rate differed among decision methods, we analyzed the episode outcomes of grain boundary navigation with Cosmos-Reason2 32B (Fig.~\ref{fig:failure}a). 
The grain boundary task was more difficult than the W dopant task, and Cosmos-Reason2 32B achieved the highest success rate in this task. 
We classified the episode outcomes into five categories: navigation success, false detection, missed detection, step limit reached, and no valid answer.
Navigation success refers to episodes that ended with the model answering \texttt{inside} at the highest magnification when the target was inside the green square.
False detection refers to episodes that ended with the model answering \texttt{inside} at the highest magnification although the target was not inside the green square.
Missed detection refers to episodes that ended with the model answering \texttt{none} at the lowest magnification although the target was within the FOV.
Step limit reached refers to episodes that did not end within 60 steps.
No valid answer refers to episodes that ended because the model did not return an answer in the specified format, which occurred only with \SO{} and \RSO{}.

The proportion of navigation success was 43\% for \TD{}, compared with 19\% for \SO{} and 2\% for \RSO{}. 
For \TD{}, the failures consisted of missed detection (36\%), step limit reached (19\%), and false detection (2\%).
For \SO{}, 61\% of all episodes ended with missed detection, and the remaining failures were step limit reached (19\%) and false detection (1\%).
For \RSO{}, 66\% of the episodes ended with false detection, and another 32\% with no valid answer.

Figure~\ref{fig:failure}b shows an example in which \TD{} and \SO{} navigated correctly, whereas \RSO{} resulted in false detection.
In this example, the green square was at the center of the initial FOV (white dashed line), and the grain boundary (light blue dashed line) was located to its right.
\TD{} repeatedly answered \texttt{right}, moving the FOV to the right, and answered \texttt{inside} at the position where the grain boundary entered the green square. 
Because \texttt{inside} was also answered at the highest magnification, navigation succeeded in 13 steps. 
\SO{} similarly moved the FOV to the right and succeeded in 13 steps. 
In contrast, \RSO{} answered that the target was inside the green square at the first decision and terminated navigation after one step. 
At this point, the grain boundary was outside the green square, and this episode was therefore classified as false detection.
Its reasoning trace stated that the grain boundary passed through the green square, although the boundary was located to the right of the square (Supplementary Note~\ref*{supp:note-reasoning}).
Unlike an incorrect directional decision, which can be corrected at the next step, an incorrect termination decision ends navigation without any opportunity for correction.

These results show that the decision method strongly affects the success rate of closed-loop navigation by changing the proportions of missed detection and false detection.
This suggests that \TD{} increases the success rate by producing fewer false detections than \RSO{} and fewer missed detections than \SO{}.
\TD{} thus had not only the shortest decision latency but also failures that could be noticed.
Most of its failures were missed detection or step limit reached, which end with the target not found, and few were false detections, which end with an incorrect report that the target was found.

\subsection{Effects of observation noise and target-description variability}
\label{sec:variability}

In our system, the VLM receives two inputs at each step: a STEM image and a natural-language target description.
In Secs.~\ref{sec:comparison} and \ref{sec:failure}, we used ideal noise-free simulated images and a single predefined target description. 
In actual observations, however, both inputs vary: image quality depends on the observation conditions, and the wording of the target description differs among operators.
VLM-based navigation must remain robust to such variability if it is to be used in diverse observations without task-specific training.
In this section, we therefore focused on \TD{} with Cosmos-Reason2 32B and evaluated how observation noise as image-side variability and changes in the target description as language-side variability affect the navigation success rate.

\begin{figure}[tbp]
  \centering
  \includegraphics[width=\linewidth]{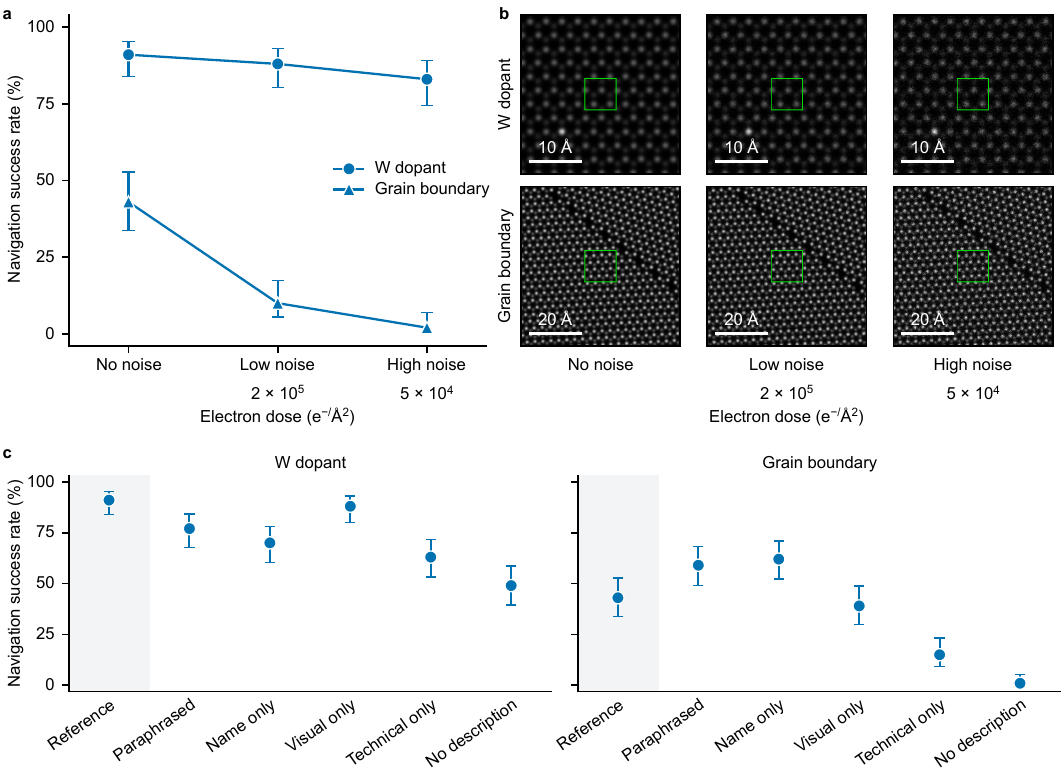}
  \caption{
  \textbf{Robustness of the VLM to noise and target descriptions.} 
  The navigation success rates in \textbf{a} and \textbf{c} were obtained by closed-loop navigation using \TD{} with Cosmos-Reason2 32B. 
  Each condition consisted of 100 episodes, and error bars indicate 95\% Wilson confidence intervals. 
  \textbf{a}, Navigation success rate when Poisson noise (shot noise) was added to the images at electron doses of $2 \times 10^{5}$~\dose{} (low noise) and $5 \times 10^{4}$~\dose{} (high noise).
  Circles and triangles indicate W dopant and grain boundary navigation, respectively. 
  \textbf{b}, Example images of the W dopant (top) and the grain boundary (bottom) at each electron dose. 
  \textbf{c}, Navigation success rate with different target descriptions (no noise) for W dopant (left) and grain boundary (right) navigation. Reference (gray band) is the description used in the experiments in Fig.~\ref{fig:methods}; Paraphrased is a paraphrase of it; Name only gives only the target name; Visual only gives only visual features; Technical only gives only a description in technical terms; and No description gives no description.
  }
  \label{fig:robustness}
\end{figure}

We first examined the effect of shot noise as image-side variability. 
Shot noise arises because the number of detected electrons follows a Poisson distribution, and it becomes more pronounced as the electron dose decreases. 
This noise is unavoidable when observations are performed at low doses to suppress beam damage\cite{Buban2010-pd}. 
We added Poisson noise to the images at electron doses of $2 \times 10^{5}$ and $5 \times 10^{4}$~\dose{} (Fig.~\ref{fig:robustness}a,b), which we refer to as low and high noise, respectively.
The electron dose per unit area was the same at all magnifications. 
For the pixel size of the simulated images (0.125 Å), these doses correspond, for example, to a probe current of 30 pA with dwell times of about 17 and 4 µs per pixel, respectively.
In the W dopant task, the success rate remained high even under high noise (83\%, compared with 91\% without noise).
In the grain boundary task, by contrast, the success rate dropped from 43\% without noise to 10\% under low noise and 2\% under high noise.
These results suggest that the W dopant remained identifiable under noise because it appears as an isolated bright spot, whereas the grain boundary became difficult to identify because it appears only as a change in lattice orientation.

Next, as language-side variability, we evaluated the success rate when the target description was changed (Fig.~\ref{fig:robustness}c). 
We compared the following six target descriptions (Supplementary Table~\ref*{supp:tab-descriptions}): (1) the description used in the preceding experiments (Reference), (2) a paraphrase of it (Paraphrased), (3) the target name only (Name only), (4) visual features only (Visual only), (5) technical terms only (Technical only), and (6) no description (No description), which served as a control to confirm that the VLM used the description.
In both tasks, the success rate was lowest with Technical only among the five descriptions (63\% for the W dopant and 15\% for the grain boundary).
These results suggest that the VLM readily relates plain expressions such as ``much brighter'' and ``grain boundary'' to image features, whereas it has difficulty relating crystallographic terms such as ``W\_Mo'' and ``$\Sigma$7 tilt''.
For the grain boundary, Paraphrased and Name only even exceeded Reference (59\% and 62\% vs.\ 43\%), indicating that the success rates for the grain boundary in Secs.~\ref{sec:comparison} and \ref{sec:failure} do not represent the upper limit achievable with a better description.
Without a description, the success rate dropped to 49\% for the W dopant and to nearly zero for the grain boundary, confirming that the VLM relied on the description in its decisions.

These results show that closed-loop navigation by \TD{} with a pretrained VLM can be applied to different targets solely through natural-language specification, without task-specific training for each target. 
However, its robustness to input variability depended on the target: the grain boundary was more affected than the W dopant by both noise and the wording of the description.

\section{Discussion}

This study shows that directly obtaining the probabilities of candidate labels from a pretrained VLM can be used for low-latency decision-making in atomic-scale closed-loop navigation.
With \TD{}, all four VLMs made decisions in 0.13--0.21~s, the shortest decision latency among the three decision methods.
Although label probabilities do not always agree with the answers generated as text\cite{Wang2024-vi}, \TD{} achieved the highest navigation success rate among the three decision methods for all models in the W dopant task and for Cosmos-Reason2 32B, the best-performing model, in the grain boundary task.
In contrast, \RSO{}, which generates an explicit reasoning trace, took 21--397 times longer per decision than \TD{}, and its navigation success rate was generally lower.
Thus, for the rapid iterative decisions required in closed-loop navigation, explicit reasoning was generally a drawback rather than a benefit.
Furthermore, changing only the natural-language target specification allowed the same VLM to navigate to two different targets without task-specific training.
These results suggest that pretrained VLMs can be used as modules responsible for iterative decision-making based on observed images.

On a real instrument, decision speed matters for more than efficiency.
Acquiring an atomic-resolution HAADF-STEM image takes several seconds per frame\cite{Goodge2020-ic, Jones2015-ca}.
The decision latency of \TD{} (0.13--0.21~s) is shorter than this, so decision-making would not become a bottleneck for the overall navigation.
In contrast, the decision latency of \RSO{} (2.7--85~s) can exceed the image acquisition time.
Because specimen drift continues while the system waits for a decision\cite{Sang2014-wn, Jones2015-ca}, the slower the decision, the larger the discrepancy between the image on which the decision was based and the actual FOV.
Therefore, the disadvantage of decisions with reasoning is expected to be larger on a real instrument than in simulation.
The decision latency can also be shortened by reducing the model size, but this generally lowers the success rate, as also observed here (Sec.~\ref{sec:comparison}).
\TD{}, in contrast, shortens the decision latency by changing how the decision is obtained rather than the model size, so a large model can still be used.

In closed-loop navigation, the most harmful type of failure is false detection, which ends navigation without the failure being noticed.
An error in the moving direction can be corrected at the next step, but an error in the termination decision ends navigation.
Missed detection ends navigation with the report that the target was not found, so the system can retry in another region or defer to an operator.
False detection instead reports that the target was found, so subsequent analysis would proceed at a wrong position.
In grain boundary navigation with Cosmos-Reason2 32B, 66\% of the episodes with \RSO{} ended with false detection, whereas the failures of \TD{} were mainly missed detection (36\%) and step limit reached (19\%), both of which end with the target not found and are therefore evident.
This result suggests that the value of a decision method in closed-loop navigation depends not only on the accuracy of a single decision but also on whether its failures can be noticed.
Missed detection can be noticed, but it still lowers the success rate and was the main cause of failure for \TD{}, so reducing it remains an important challenge.

Another feature of our approach is that the VLM is connected to microscope operations through a small set of predefined labels.
Because \TD{} obtains the probabilities of all candidate labels at every decision, its output is always one of the predefined labels.
Accordingly, \TD{} returned a valid answer at every decision, whereas, in grain boundary navigation with Cosmos-Reason2 32B, \RSO{} failed to return a valid answer in 32\% of the episodes.
The output of the VLM is then passed to the fixed navigation policy, which converts it into operations such as FOV translation and magnification changes.
Because this design separates image interpretation from the rules for microscope operation, constraints such as the travel range of the specimen stage are enforced by the rules rather than left to the VLM.

In previous autonomous EM systems, the target had to be defined for each task, either by training a model on labeled images or by designing a quantity to be maximized\cite{Creange2022-mh,Addiego2025-gr,Roccapriore2022-nm}.
In contrast, pretrained VLMs can be applied to different navigation tasks by changing the natural-language target specification, without task-specific training.
Indeed, the same VLM navigated to both the W dopant and the grain boundary in this study.
However, the navigation success rate varied with the recognition performance of the model, observation noise, and the content of the target description.
In particular, the change in performance under noise differed greatly between the targets.
For the target description, giving the name or visual features of the target in plain words was more effective than using crystallographic terms.
This suggests that, before VLM-based navigation is applied to a new type of target, its success rate should be checked with the intended description and under the expected noise level\cite{Alampara2025-bd}.

Taken together, \TD{} combines short decision latency, evident failures, and guaranteed valid output.
Methods that directly obtain label probabilities from pretrained VLMs are therefore promising for iterative image-based decisions in autonomous electron microscopy.
Closed-loop STEM requires both observation planning, which determines what to observe, and sequential decision-making, which selects the next operation from each acquired image.
These two roles could be assigned to different methods: observation planning, which requires deliberation, to a model with explicit reasoning, and sequential decision-making to a fast method without reasoning, such as \TD{}.
Furthermore, because \TD{} provides label probabilities, it is also possible to delegate only low-confidence decisions to a higher-level model or a human\cite{Li2026-ve}.
In addition, the candidate labels and the navigation policy used in this study do not depend on the target or the material, so the same framework is expected to be applicable to other targets and materials.
With a different set of labels and policy, the framework could also be extended to other sequential decisions, such as selecting positions for spectroscopic measurements by energy-dispersive X-ray spectroscopy (EDS) or EELS and adjusting observation conditions.

However, our evaluation was performed in a simulated environment, and we did not examine the effects of factors present in actual microscopes, such as specimen drift, instrument response delays, and beam damage\cite{Khan2023-eh}. 
In addition, the navigation targets were limited to two types of target in monolayer \ce{MoS2}. 
On the other hand, in a simulated environment, the true position of the target is known, and multiple decision methods can be compared repeatedly from identical initial conditions. 
On a real instrument, beam damage makes it difficult to repeat navigation in the same region under the same conditions. 
Our evaluation can thus be regarded as an idealized task for isolating the properties of the decision methods themselves. 
Future work should verify the practicality of this approach through closed-loop navigation on a real instrument, apply it to different materials and observation conditions, and examine whether image preprocessing, such as intensity inversion and segmentation-based highlighting, reduces missed detection, the main cause of failure for \TD{}, by improving target recognition.

\section{Methods}

\subsection{Simulated STEM environments}

We constructed two simulated HAADF-STEM environments, one targeting a W dopant and the other targeting a grain boundary, both in monolayer \ce{MoS2}.
HAADF-STEM images were generated from the specimen structures using abTEM (v1.0.10)\cite{Madsen2021-bu}.
Monolayer 2H-\ce{MoS2} ($a = 3.16$~\AA) was built with ASE\cite{Hjorth-Larsen2017-us}, and two specimens were created from it: one in which a single Mo atom was substituted with W, and one containing a $\Sigma 7$ symmetric tilt grain boundary (misorientation of $21.79^{\circ}$). 
The grain boundary was constructed using the cut-and-mirror method, and no structural relaxation was performed. 
We used an accelerating voltage of 80~kV, a convergence semi-angle of 25~mrad, and detector collection angles of 70--200~mrad, with an aberration-free probe at exact focus ($\Delta f = 0$, $C_s = 0$).
The projected potential was calculated using the Lobato parametrization\cite{Lobato2014-uu} with a sampling of 0.06~\AA{} and a slice thickness of 1~\AA, and the frozen phonon method was applied with four configurations (standard deviations of atomic displacements of 0.06~\AA{} for Mo and W and 0.08~\AA{} for S).
The probe was scanned with a step of 0.35~\AA, and the images were interpolated to 0.125~\AA/pixel and convolved with a Gaussian of $\sigma = 0.3$~\AA{} to account for the finite effective source size.
In each episode, the W dopant image was randomly translated, and the grain boundary image was randomly rotated and flipped.

The virtual microscope we built allowed the FOV center to be translated and the magnification to be switched among three levels. 
The FOV widths were approximately 32, 64, and 96~\AA{} for the W dopant task and approximately 64, 96, and 128~\AA{} for the grain boundary task. 
Each observed image was $512 \times 512$ pixels.
A green square with a side length of 0.2 times the FOV width was overlaid at the center of each observed image.
For the noise experiments, Poisson noise was added to the simulated image at its original pixel size (0.125~\AA) before resizing, so that the dose per unit area was the same at all magnifications.
The number of detected electrons in each pixel was drawn from a Poisson distribution with a mean of $IN$, where $I$ is the simulated intensity, given as the fraction of incident electrons reaching the detector, and $N = D \times (0.125~\text{\AA})^{2}$ is the number of incident electrons per pixel at a dose $D$.

\subsection{VLM decision methods}

Using four VLMs (Cosmos-Reason2, Qwen3.8, Gemma 4, and Nemotron 3 Nano Omni), we compared three decision methods: \TD{}, \SO{}, and \RSO{}. 
For the model-size comparison, Cosmos-Reason2 2B and 8B were also used (Supplementary Fig.~\ref*{supp:fig-size}).
The Hugging Face repositories and the checkpoints used for each model are listed in Supplementary Table~\ref*{supp:tab-models}. 
All methods were given the same inputs: a STEM image, a natural-language target description, and a question about the relative position of the target. 
The answer was obtained as one of six labels: \texttt{left}, \texttt{right}, \texttt{up}, \texttt{down}, \texttt{inside}, and \texttt{none}.

In \TD{}, no answer was generated autoregressively. 
Instead, the next-token log-probabilities of the six candidate labels were obtained directly, and the label with the highest normalized probability was selected.
All six labels were single tokens for all the models used. 
In \SO{}, the answer format was constrained by a JSON schema, and the label was obtained through autoregressive generation. 
In \RSO{}, the model first generated an explicit reasoning trace, enabled by model-specific settings or instructions in the prompt (Supplementary Note~\ref*{supp:note-prompts}), and then generated the same JSON answer as in \SO{}.

Inference was performed with vLLM\cite{Kwon2023-gg} on a single NVIDIA RTX PRO 6000 Blackwell GPU, and the sampling parameters for generation followed the default values of each model.
Cosmos-Reason2 was run in BF16, and the other three models were run with NVIDIA's NVFP4 quantized weights.
The decision latency was defined as the mean time per decision, measured by sending requests one at a time.

When \SO{} or \RSO{} did not return a valid answer, the episode was terminated without a re-query and recorded as no valid answer. 
In most of these episodes, the output reached the maximum of 8192 output tokens before an answer was produced.
\TD{} always returned a valid answer, because the probabilities of all six labels are obtained at every decision.

\subsection{Closed-loop navigation policy}

To convert the VLM decisions into microscope operations, we implemented a fixed navigation policy shared by all decision methods. 
When the target was judged to be inside the green square, navigation was terminated at the highest magnification; at any other magnification, the magnification was increased by one level. 
When the target was judged to be absent from the FOV, the magnification was decreased by one level; at the lowest magnification, navigation was terminated.
An episode was also terminated when no valid answer was obtained or the step limit of 60 steps was reached.

For directional answers, the movement vector was computed from the probability differences between \texttt{left} and \texttt{right} and between \texttt{up} and \texttt{down} as follows:
\begin{equation}
  \mathbf{v} =
  \begin{pmatrix}
    P(\text{right}) - P(\text{left}) \\
    P(\text{down}) - P(\text{up})
  \end{pmatrix}.
  \label{eq:move}
\end{equation}
For the probabilities in this vector, \TD{} used the predicted probabilities of the labels, whereas \SO{} and \RSO{} used the one-hot probabilities of the selected label. 
The movement direction was determined by normalizing this vector, and the movement distance was set to 1/16 of the FOV width. 

Directions that would exceed the travel range of the specimen stage were excluded from the candidate labels.
To prevent repeated zooming in and out at the same position, \texttt{inside} was also excluded at a position where the target had been lost immediately after a magnification increase. 
In \TD{}, the probabilities of the remaining labels were renormalized; in \SO{} and \RSO{}, the model was re-queried only once when it returned an excluded label.

\subsection{Evaluation protocol}

Closed-loop navigation was evaluated for each combination of task, model, and decision method over 100 episodes, which combined 50 random seeds with two starting conditions (target inside or outside the initial FOV). 
When the target was outside the initial FOV, the starting position was chosen such that the target came into the FOV when the magnification was decreased to the lowest level at that position.
Each episode started at the highest magnification, and the same initial conditions were used for all decision methods. 
The ground-truth label was \texttt{inside} when the center of the W atom lay within the green square in the W dopant task and when the grain boundary line intersected the green square in the grain boundary task.
Navigation was counted as successful only when the model answered \texttt{inside} at the highest magnification and the ground-truth label was also \texttt{inside}.
The navigation success rate was defined as the fraction of successful episodes among all episodes, and 95\% confidence intervals were calculated using the Wilson score method\cite{Wilson1927-dc}.

\section*{Data availability}

The simulated HAADF-STEM images and the atomic model of the grain boundary used in this study are available at \url{https://github.com/shu-ozawa/vlm-stem-navigation}, together with the simulation parameters.

\section*{Code availability}

The code used for the STEM image simulations and the closed-loop navigation in this study is available at \url{https://github.com/shu-ozawa/vlm-stem-navigation}.

\section*{Acknowledgements}

This study was supported by the Japan Science and Technology Agency (JST) (Nos. JPMJAX24DB and JPMJBS2418) and the Ministry of Education, Culture, Sports, Science and Technology (MEXT), Japan (Nos. 24H00042 and 26K01205). 
This study was also supported by the MEXT Supporting Pioneering Research through AI for 1000 Discovery challenges Program (SPReAD) (Grant No. JPMXP1726304204).

\bibliographystyle{unsrtnat}
\bibliography{references}

\begin{thebibliography}{57}
\providecommand{\natexlab}[1]{#1}
\providecommand{\url}[1]{\texttt{#1}}
\expandafter\ifx\csname urlstyle\endcsname\relax
  \providecommand{\doi}[1]{doi: #1}\else
  \providecommand{\doi}{doi: \begingroup \urlstyle{rm}\Url}\fi

\bibitem[Susi(2022)]{Susi2022-oh}
Toma Susi.
\newblock Identifying and manipulating single atoms with scanning transmission
  electron microscopy.
\newblock \emph{Chem. Commun. (Camb.)}, 58\penalty0 (88):\penalty0
  12274--12285, 3~November 2022.

\bibitem[Ophus(2023)]{Ophus2023-jc}
Colin Ophus.
\newblock Quantitative scanning transmission electron microscopy for materials
  science: Imaging, diffraction, spectroscopy, and tomography.
\newblock \emph{Annu. Rev. Mater. Res.}, 53\penalty0 (1):\penalty0 105--141,
  3~July 2023.

\bibitem[Kalinin et~al.(2021)Kalinin, Ziatdinov, Hinkle, Jesse, Ghosh, Kelley,
  Lupini, Sumpter, and Vasudevan]{Kalinin2021-ko}
Sergei~V Kalinin, Maxim Ziatdinov, Jacob Hinkle, Stephen Jesse, Ayana Ghosh,
  Kyle~P Kelley, Andrew~R Lupini, Bobby~G Sumpter, and Rama~K Vasudevan.
\newblock Automated and autonomous experiments in electron and scanning probe
  microscopy.
\newblock \emph{ACS Nano}, 15\penalty0 (8):\penalty0 12604--12627, 24~August
  2021.

\bibitem[Spurgeon et~al.(2021)Spurgeon, Ophus, Jones, Petford-Long, Kalinin,
  Olszta, Dunin-Borkowski, Salmon, Hattar, Yang, Sharma, Du, Chiaramonti,
  Zheng, Buck, Kovarik, Penn, Li, Zhang, Murayama, and Taheri]{Spurgeon2021-rx}
Steven~R Spurgeon, Colin Ophus, Lewys Jones, Amanda Petford-Long, Sergei~V
  Kalinin, Matthew~J Olszta, Rafal~E Dunin-Borkowski, Norman Salmon, Khalid
  Hattar, Wei-Chang~D Yang, Renu Sharma, Yingge Du, Ann Chiaramonti, Haimei
  Zheng, Edgar~C Buck, Libor Kovarik, R~Lee Penn, Dongsheng Li, Xin Zhang,
  Mitsuhiro Murayama, and Mitra~L Taheri.
\newblock Towards data-driven next-generation transmission electron microscopy.
\newblock \emph{Nat. Mater.}, 20\penalty0 (3):\penalty0 274--279, March 2021.

\bibitem[Kalinin et~al.(2023)Kalinin, Mukherjee, Roccapriore, Blaiszik, Ghosh,
  Ziatdinov, Al-Najjar, Doty, Akers, Rao, Agar, and Spurgeon]{Kalinin2023-ys}
Sergei~V Kalinin, Debangshu Mukherjee, Kevin Roccapriore, Benjamin~J Blaiszik,
  Ayana Ghosh, Maxim~A Ziatdinov, Anees Al-Najjar, Christina Doty, Sarah Akers,
  Nageswara~S Rao, Joshua~C Agar, and Steven~R Spurgeon.
\newblock Machine learning for automated experimentation in scanning
  transmission electron microscopy.
\newblock \emph{Npj Comput. Mater.}, 9\penalty0 (1):\penalty0 227, 20~December
  2023.

\bibitem[Jamali et~al.(2026)Jamali, Aghazadeh, and Kacher]{Jamali2026-bf}
Vida Jamali, Amirali Aghazadeh, and Josh Kacher.
\newblock Thinking microscopes: agentic {AI} and the future of electron
  microscopy.
\newblock \emph{Npj Comput. Mater.}, 12\penalty0 (1):\penalty0 149, 10~April
  2026.

\bibitem[Burger et~al.(2020)Burger, Maffettone, Gusev, Aitchison, Bai, Wang,
  Li, Alston, Li, Clowes, Rankin, Harris, Sprick, and Cooper]{Burger2020-zb}
Benjamin Burger, Phillip~M Maffettone, Vladimir~V Gusev, Catherine~M Aitchison,
  Yang Bai, Xiaoyan Wang, Xiaobo Li, Ben~M Alston, Buyi Li, Rob Clowes, Nicola
  Rankin, Brandon Harris, Reiner~Sebastian Sprick, and Andrew~I Cooper.
\newblock A mobile robotic chemist.
\newblock \emph{Nature}, 583\penalty0 (7815):\penalty0 237--241, 8~July 2020.

\bibitem[MacLeod et~al.(2020)MacLeod, Parlane, Morrissey, Häse, Roch,
  Dettelbach, Moreira, Yunker, Rooney, Deeth, Lai, Ng, Situ, Zhang, Elliott,
  Haley, Dvorak, Aspuru-Guzik, Hein, and Berlinguette]{MacLeod2020-zs}
B~P MacLeod, F~G~L Parlane, T~D Morrissey, F~Häse, L~M Roch, K~E Dettelbach,
  R~Moreira, L~P~E Yunker, M~B Rooney, J~R Deeth, V~Lai, G~J Ng, H~Situ, R~H
  Zhang, M~S Elliott, T~H Haley, D~J Dvorak, A~Aspuru-Guzik, J~E Hein, and C~P
  Berlinguette.
\newblock Self-driving laboratory for accelerated discovery of thin-film
  materials.
\newblock \emph{Sci. Adv.}, 6\penalty0 (20):\penalty0 eaaz8867, May 2020.

\bibitem[Tom et~al.(2024)Tom, Schmid, Baird, Cao, Darvish, Hao, Lo,
  Pablo-García, Rajaonson, Skreta, Yoshikawa, Corapi, Akkoc, Strieth-Kalthoff,
  Seifrid, and Aspuru-Guzik]{Tom2024-yh}
Gary Tom, Stefan~P Schmid, Sterling~G Baird, Yang Cao, Kourosh Darvish, Han
  Hao, Stanley Lo, Sergio Pablo-García, Ella~M Rajaonson, Marta Skreta, Naruki
  Yoshikawa, Samantha Corapi, Gun~Deniz Akkoc, Felix Strieth-Kalthoff, Martin
  Seifrid, and Alán Aspuru-Guzik.
\newblock Self-driving laboratories for chemistry and materials science.
\newblock \emph{Chem. Rev.}, 124\penalty0 (16):\penalty0 9633--9732, 28~August
  2024.

\bibitem[Szymanski et~al.(2023)Szymanski, Rendy, Fei, Kumar, He, Milsted,
  McDermott, Gallant, Cubuk, Merchant, Kim, Jain, Bartel, Persson, Zeng, and
  Ceder]{Szymanski2023-al}
Nathan~J Szymanski, Bernardus Rendy, Yuxing Fei, Rishi~E Kumar, Tanjin He,
  David Milsted, Matthew~J McDermott, Max Gallant, Ekin~Dogus Cubuk, Amil
  Merchant, Haegyeom Kim, Anubhav Jain, Christopher~J Bartel, Kristin Persson,
  Yan Zeng, and Gerbrand Ceder.
\newblock An autonomous laboratory for the accelerated synthesis of inorganic
  materials.
\newblock \emph{Nature}, 624\penalty0 (7990):\penalty0 86--91, November 2023.

\bibitem[Takahara et~al.(2026)Takahara, Nishio, Aiba, Kobayashi, Nakajima,
  Hitosugi, and Mizoguchi]{Takahara2026-hy}
Izumi Takahara, Kazunori Nishio, Akira Aiba, Shigeru Kobayashi, Takao Nakajima,
  Taro Hitosugi, and Teruyasu Mizoguchi.
\newblock Hypothesis-driven autonomous materials synthesis with multimodal
  {LLM} agents.
\newblock \emph{arXiv [cond-mat.mtrl-sci]}, 2026.

\bibitem[Creange et~al.(2022)Creange, Dyck, Vasudevan, Ziatdinov, and
  Kalinin]{Creange2022-mh}
Nicole Creange, Ondrej Dyck, Rama~K Vasudevan, Maxim Ziatdinov, and Sergei~V
  Kalinin.
\newblock Towards automating structural discovery in scanning transmission
  electron microscopy.
\newblock \emph{Mach. Learn. Sci. Technol.}, 3\penalty0 (1):\penalty0 015024,
  1~March 2022.

\bibitem[Addiego et~al.(2025)Addiego, Huang, Han, and Zhu]{Addiego2025-gr}
Christopher Addiego, Ying Huang, Xiuhong Han, and Jiangtao Zhu.
\newblock A flexible system for fully autonomous {STEM} imaging.
\newblock \emph{Microscopy and Microanalysis}, 31\penalty0
  (Supplement\_1):\penalty0 ozaf048.1089, 2025.

\bibitem[Roccapriore et~al.(2022)Roccapriore, Kalinin, and
  Ziatdinov]{Roccapriore2022-nm}
Kevin~M Roccapriore, Sergei~V Kalinin, and Maxim Ziatdinov.
\newblock Physics discovery in nanoplasmonic systems via autonomous experiments
  in scanning transmission electron microscopy.
\newblock \emph{Adv. Sci. (Weinh.)}, 9\penalty0 (36):\penalty0 e2203422,
  1~December 2022.

\bibitem[Ziatdinov et~al.(2017)Ziatdinov, Dyck, Maksov, Li, Sang, Xiao, Unocic,
  Vasudevan, Jesse, and Kalinin]{Ziatdinov2017-te}
Maxim Ziatdinov, Ondrej Dyck, Artem Maksov, Xufan Li, Xiahan Sang, Kai Xiao,
  Raymond~R Unocic, Rama Vasudevan, Stephen Jesse, and Sergei~V Kalinin.
\newblock Deep learning of atomically resolved scanning transmission electron
  microscopy images: Chemical identification and tracking local
  transformations.
\newblock \emph{ACS Nano}, 11\penalty0 (12):\penalty0 12742--12752, 26~December
  2017.

\bibitem[Maksov et~al.(2019)Maksov, Dyck, Wang, Xiao, Geohegan, Sumpter,
  Vasudevan, Jesse, Kalinin, and Ziatdinov]{Maksov2019-ff}
Artem Maksov, Ondrej Dyck, Kai Wang, Kai Xiao, David~B Geohegan, Bobby~G
  Sumpter, Rama~K Vasudevan, Stephen Jesse, Sergei~V Kalinin, and Maxim
  Ziatdinov.
\newblock Deep learning analysis of defect and phase evolution during electron
  beam-induced transformations in {WS2}.
\newblock \emph{Npj Comput. Mater.}, 5\penalty0 (1):\penalty0 12, 1~February
  2019.

\bibitem[Yang et~al.(2021)Yang, Choi, Cho, Agyapong-Fordjour, Park, Yun, Kim,
  Han, Lee, Kim, and Kim]{Yang2021-ts}
Sang-Hyeok Yang, Wooseon Choi, Byeong~Wook Cho, Frederick Osei-Tutu
  Agyapong-Fordjour, Sehwan Park, Seok~Joon Yun, Hyung-Jin Kim, Young-Kyu Han,
  Young~Hee Lee, Ki~Kang Kim, and Young-Min Kim.
\newblock Deep learning-assisted quantification of atomic dopants and defects
  in {2D} materials.
\newblock \emph{Adv. Sci. (Weinh.)}, 8\penalty0 (16):\penalty0 e2101099,
  1~August 2021.

\bibitem[Akers et~al.(2021)Akers, Kautz, Trevino-Gavito, Olszta, Matthews,
  Wang, Du, and Spurgeon]{Akers2021-xh}
Sarah Akers, Elizabeth Kautz, Andrea Trevino-Gavito, Matthew Olszta, Bethany~E
  Matthews, Le~Wang, Yingge Du, and Steven~R Spurgeon.
\newblock Rapid and flexible segmentation of electron microscopy data using
  few-shot machine learning.
\newblock \emph{Npj Comput. Mater.}, 7\penalty0 (1):\penalty0 187, 17~November
  2021.

\bibitem[Yang et~al.(2025)Yang, Yin, Jiang, Hu, Zhu, Hu, Kumar, Holmes, Wang,
  Zhai, Rong, Zhu, Zhang, Yin, Cao, Tang, Franklin, Kong, Gong, Ren, and
  Wang]{Yang2025-nj}
Jingyun Yang, Ruoyan~Avery Yin, Chi Jiang, Yuepeng Hu, Xiaokai Zhu, Xingjian
  Hu, Sutharsika Kumar, Samantha~K Holmes, Xiao Wang, Xiaohua Zhai, Keran Rong,
  Yunyue Zhu, Tianyi Zhang, Zongyou Yin, Yuan Cao, Haoning Tang, Aaron~D
  Franklin, Jing Kong, Neil~Zhenqiang Gong, Zhichu Ren, and Haozhe Wang.
\newblock Zero-shot autonomous microscopy for scalable and intelligent
  characterization of {2D} materials.
\newblock \emph{ACS Nano}, 19\penalty0 (40):\penalty0 35493--35502, 14~October
  2025.

\bibitem[Wall et~al.(2026)Wall, Pattison, Barnard, Ribet, and
  Ercius]{Wall2026-uu}
Morgan~K Wall, Alexander~J Pattison, Edward~S Barnard, Stephanie~M Ribet, and
  Peter Ercius.
\newblock {TEM} agent: enhancing transmission electron microscopy with modern
  {AI} tools.
\newblock \emph{Npj Comput. Mater.}, 12\penalty0 (1):\penalty0 316, 10~June
  2026.

\bibitem[Du et~al.(2026)Du, Luo, Banerjee, Wojcik, Popovic, and
  Cherukara]{Du2026-po}
Ming Du, Yanqi Luo, Srutarshi Banerjee, Michael Wojcik, Jelena Popovic, and
  Mathew~J Cherukara.
\newblock Experiment automation agents: Automating materials characterization
  with vision language model agents.
\newblock \emph{Npj Comput. Mater.}, pages 1--12, 31~July 2026.

\bibitem[Chen et~al.(2026)Chen, Yuan, and You]{Chen2026-pr}
Guangyao Chen, Wenhao Yuan, and Fengqi You.
\newblock Bridging electron microscopy and materials analysis with an
  autonomous agentic platform.
\newblock \emph{Sci. Adv.}, 12\penalty0 (14):\penalty0 eaed0583, 3~April 2026.

\bibitem[Polat et~al.(2026)Polat, Serpedin, Kurban, and Kurban]{Polat2026-bi}
Can Polat, Erchin Serpedin, Mustafa Kurban, and Hasan Kurban.
\newblock {STEMGym}: Benchmarking sequential decision-making under dose budgets
  in autonomous electron microscopy.
\newblock \emph{arXiv [cs.LG]}, 28~June 2026.

\bibitem[Egerton et~al.(2004)Egerton, Li, and Malac]{Egerton2004-zt}
R~F Egerton, P~Li, and M~Malac.
\newblock Radiation damage in the {TEM} and {SEM}.
\newblock \emph{Micron}, 35\penalty0 (6):\penalty0 399--409, 1~August 2004.

\bibitem[Komsa et~al.(2012)Komsa, Kotakoski, Kurasch, Lehtinen, Kaiser, and
  Krasheninnikov]{Komsa2012-ml}
Hannu-Pekka Komsa, Jani Kotakoski, Simon Kurasch, Ossi Lehtinen, Ute Kaiser,
  and Arkady~V Krasheninnikov.
\newblock Two-dimensional transition metal dichalcogenides under electron
  irradiation: defect production and doping.
\newblock \emph{Phys. Rev. Lett.}, 109\penalty0 (3):\penalty0 035503, 20~July
  2012.

\bibitem[Alayrac et~al.(2022)Alayrac, Donahue, Luc, Miech, Barr, Hasson, Lenc,
  Mensch, Millican, Reynolds, Ring, Rutherford, Cabi, Han, Gong, Samangooei,
  Monteiro, Menick, Borgeaud, Brock, Nematzadeh, Sharifzadeh, Binkowski,
  Barreira, Vinyals, Zisserman, and Simonyan]{Alayrac2022-qz}
Jean-Baptiste Alayrac, Jeff Donahue, Pauline Luc, Antoine Miech, Iain Barr,
  Yana Hasson, Karel Lenc, Arthur Mensch, Katherine Millican, Malcolm Reynolds,
  Roman Ring, Eliza Rutherford, Serkan Cabi, Tengda Han, Zhitao Gong, Sina
  Samangooei, Marianne Monteiro, Jacob~L Menick, Sebastian Borgeaud, Andy
  Brock, Aida Nematzadeh, Sahand Sharifzadeh, Mikolaj Binkowski, Ricardo
  Barreira, Oriol Vinyals, Andrew Zisserman, and Karén Simonyan.
\newblock Flamingo: A visual language model for few-shot learning.
\newblock In S~Koyejo, S~Mohamed, A~Agarwal, D~Belgrave, K~Cho, and A~Oh,
  editors, \emph{Advances in Neural Information Processing Systems 35},
  volume~35, pages 23716--23736, San Diego, California, USA, 2022. Neural
  Information Processing Systems Foundation, Inc. (NeurIPS).

\bibitem[Wang et~al.(2024)Wang, Ma, Hu, Weber-Genzel, Röttger, Kreuter, Hovy,
  and Plank]{Wang2024-vi}
Xinpeng Wang, Bolei Ma, Chengzhi Hu, Leon Weber-Genzel, Paul Röttger, Frauke
  Kreuter, Dirk Hovy, and Barbara Plank.
\newblock “my answer is c”: First-token probabilities do not match text
  answers in instruction-tuned language models.
\newblock In \emph{Findings of the Association for Computational Linguistics
  ACL 2024}, pages 7407--7416, Stroudsburg, PA, USA, 2024. Association for
  Computational Linguistics.

\bibitem[Li et~al.(2024)Li, Ge, Ge, Wang, Wang, Zhang, and Shan]{Li2024-ki}
Bohao Li, Yuying Ge, Yixiao Ge, Guangzhi Wang, Rui Wang, Ruimao Zhang, and Ying
  Shan.
\newblock {SEED}-bench: Benchmarking multimodal large language models.
\newblock In \emph{2024 IEEE/CVF Conference on Computer Vision and Pattern
  Recognition (CVPR)}, pages 13299--13308. IEEE, 16~June 2024.

\bibitem[Almeida(2026)]{Almeida2026-sy}
Diogo Almeida.
\newblock Introducing {System One} models \& {Jev}.
\newblock TypeSafe AI blog,
  \url{https://typesafe.ai/blog/introducing-system-one-models-and-jev},
  September 2026.
\newblock Accessed 2026-11-17.

\bibitem[Tang and Zheng(2026)]{Tang2026-td}
Lijuan Tang and Yuemeng Zheng.
\newblock Typed decision models: An early evidence audit and evaluation
  checklist.
\newblock \emph{arXiv [cs.CL]}, 26~September 2026.

\bibitem[Wei et~al.(2022)Wei, Wang, Schuurmans, Bosma, Ichter, Xia, Chi, Le,
  and Zhou]{Wei2022-mz}
Jason Wei, Xuezhi Wang, Dale Schuurmans, Maarten Bosma, Brian Ichter, Fei Xia,
  Ed~Chi, Quoc~V Le, and Denny Zhou.
\newblock Chain-of-thought prompting elicits reasoning in large language
  models.
\newblock In \emph{Advances in Neural Information Processing Systems 35}, pages
  24824--24837, San Diego, California, USA, 2022. Neural Information Processing
  Systems Foundation, Inc. (NeurIPS).

\bibitem[Yao et~al.(2022)Yao, Zhao, Yu, Du, Shafran, Narasimhan, and
  Cao]{Yao2022-vs}
Shunyu Yao, Jeffrey Zhao, Dian Yu, Nan Du, Izhak Shafran, Karthik Narasimhan,
  and Yuan Cao.
\newblock {ReAct}: Synergizing reasoning and acting in language models.
\newblock \emph{arXiv [cs.CL]}, 6~October 2022.

\bibitem[Zhang et~al.(2023)Zhang, Zhang, Li, Zhao, Karypis, and
  Smola]{Zhang2023-xo}
Zhuosheng Zhang, Aston Zhang, Mu~Li, Hai Zhao, George Karypis, and Alex Smola.
\newblock Multimodal chain-of-thought reasoning in language models.
\newblock \emph{arXiv [cs.CL]}, 2~February 2023.

\bibitem[Zawalski et~al.(2024)Zawalski, Chen, Pertsch, Mees, Finn, and
  Levine]{Zawalski2024-it}
Michał Zawalski, William Chen, Karl Pertsch, Oier Mees, Chelsea Finn, and
  Sergey Levine.
\newblock Robotic control via embodied chain-of-thought reasoning.
\newblock \emph{arXiv [cs.RO]}, 11~July 2024.

\bibitem[Splendiani et~al.(2010)Splendiani, Sun, Zhang, Li, Kim, Chim, Galli,
  and Wang]{Splendiani2010-ep}
Andrea Splendiani, Liang Sun, Yuanbo Zhang, Tianshu Li, Jonghwan Kim, Chi-Yung
  Chim, Giulia Galli, and Feng Wang.
\newblock Emerging photoluminescence in monolayer {MoS$_2$}.
\newblock \emph{Nano Lett.}, 10\penalty0 (4):\penalty0 1271--1275, April 2010.

\bibitem[Manzeli et~al.(2017)Manzeli, Ovchinnikov, Pasquier, Yazyev, and
  Kis]{Manzeli2017-td}
Sajedeh Manzeli, Dmitry Ovchinnikov, Diego Pasquier, Oleg~V Yazyev, and Andras
  Kis.
\newblock {2D} transition metal dichalcogenides.
\newblock \emph{Nat. Rev. Mater.}, 2\penalty0 (8):\penalty0 17033, 13~June
  2017.

\bibitem[van~der Zande et~al.(2013)van~der Zande, Huang, Chenet, Berkelbach,
  You, Lee, Heinz, Reichman, Muller, and Hone]{van-der-Zande2013-hi}
Arend~M van~der Zande, Pinshane~Y Huang, Daniel~A Chenet, Timothy~C Berkelbach,
  YuMeng You, Gwan-Hyoung Lee, Tony~F Heinz, David~R Reichman, David~A Muller,
  and James~C Hone.
\newblock Grains and grain boundaries in highly crystalline monolayer
  molybdenum disulphide.
\newblock \emph{Nat. Mater.}, 12\penalty0 (6):\penalty0 554--561, 5~June 2013.

\bibitem[Zhou et~al.(2013)Zhou, Zou, Najmaei, Liu, Shi, Kong, Lou, Ajayan,
  Yakobson, and Idrobo]{Zhou2013-ub}
Wu~Zhou, Xiaolong Zou, Sina Najmaei, Zheng Liu, Yumeng Shi, Jing Kong, Jun Lou,
  Pulickel~M Ajayan, Boris~I Yakobson, and Juan-Carlos Idrobo.
\newblock Intrinsic structural defects in monolayer molybdenum disulfide.
\newblock \emph{Nano Lett.}, 13\penalty0 (6):\penalty0 2615--2622, 12~June
  2013.

\bibitem[Krivanek et~al.(2010)Krivanek, Chisholm, Nicolosi, Pennycook, Corbin,
  Dellby, Murfitt, Own, Szilagyi, Oxley, Pantelides, and
  Pennycook]{Krivanek2010-hg}
Ondrej~L Krivanek, Matthew~F Chisholm, Valeria Nicolosi, Timothy~J Pennycook,
  George~J Corbin, Niklas Dellby, Matthew~F Murfitt, Christopher~S Own,
  Zoltan~S Szilagyi, Mark~P Oxley, Sokrates~T Pantelides, and Stephen~J
  Pennycook.
\newblock Atom-by-atom structural and chemical analysis by annular dark-field
  electron microscopy.
\newblock \emph{Nature}, 464\penalty0 (7288):\penalty0 571--574, 25~March 2010.

\bibitem[Pennycook and Jesson(1991)]{Pennycook1991-hm}
S~J Pennycook and D~E Jesson.
\newblock High-resolution {Z}-contrast imaging of crystals.
\newblock \emph{Ultramicroscopy}, 37\penalty0 (1-4):\penalty0 14--38, 1~August
  1991.

\bibitem[Dumcenco et~al.(2013)Dumcenco, Kobayashi, Liu, Huang, and
  Suenaga]{Dumcenco2013-vq}
Dumitru~O Dumcenco, Haruka Kobayashi, Zheng Liu, Ying-Sheng Huang, and Kazu
  Suenaga.
\newblock Visualization and quantification of transition metal atomic mixing in
  {Mo$_{1-x}$W$_x$S$_2$} single layers.
\newblock \emph{Nat. Commun.}, 4:\penalty0 1351, 15~January 2013.

\bibitem[Najmaei et~al.(2013)Najmaei, Liu, Zhou, Zou, Shi, Lei, Yakobson,
  Idrobo, Ajayan, and Lou]{Najmaei2013-qj}
Sina Najmaei, Zheng Liu, Wu~Zhou, Xiaolong Zou, Gang Shi, Sidong Lei, Boris~I
  Yakobson, Juan-Carlos Idrobo, Pulickel~M Ajayan, and Jun Lou.
\newblock Vapour phase growth and grain boundary structure of molybdenum
  disulphide atomic layers.
\newblock \emph{Nat. Mater.}, 12\penalty0 (8):\penalty0 754--759, 9~August
  2013.

\bibitem[Yazyev and Louie(2010)]{Yazyev2010-hw}
Oleg~V Yazyev and Steven~G Louie.
\newblock Topological defects in graphene: Dislocations and grain boundaries.
\newblock \emph{Phys. Rev. B Condens. Matter Mater. Phys.}, 81\penalty0
  (19):\penalty0 195420, 14~May 2010.

\bibitem[Shtedritski et~al.(2023)Shtedritski, Rupprecht, and
  Vedaldi]{Shtedritski2023-wg}
Aleksandar Shtedritski, Christian Rupprecht, and Andrea Vedaldi.
\newblock What does {CLIP} know about a red circle? visual prompt engineering
  for {VLMs}.
\newblock In \emph{2023 IEEE/CVF International Conference on Computer Vision
  (ICCV)}, pages 11953--11963. IEEE, 1~October 2023.

\bibitem[{NVIDIA}(2026)]{NVIDIA2026CosmosReason2}
{NVIDIA}.
\newblock {Cosmos-Reason2-32B}.
\newblock \url{https://huggingface.co/nvidia/Cosmos-Reason2-32B}, 2026.
\newblock Accessed: 2026-11-09.

\bibitem[Buban et~al.(2010)Buban, Ramasse, Gipson, Browning, and
  Stahlberg]{Buban2010-pd}
James~P Buban, Quentin Ramasse, Bryant Gipson, Nigel~D Browning, and Henning
  Stahlberg.
\newblock High-resolution low-dose scanning transmission electron microscopy.
\newblock \emph{J. Electron Microsc. (Tokyo)}, 59\penalty0 (2):\penalty0
  103--112, 2010.

\bibitem[Goodge et~al.(2020)Goodge, Bianco, Schnitzer, Zandbergen, and
  Kourkoutis]{Goodge2020-ic}
Berit~H Goodge, Elisabeth Bianco, Noah Schnitzer, Henny~W Zandbergen, and
  Lena~F Kourkoutis.
\newblock Atomic-resolution cryo-{STEM} across continuously variable
  temperatures.
\newblock \emph{Microsc. Microanal.}, 26\penalty0 (3):\penalty0 439--446,
  1~June 2020.

\bibitem[Jones et~al.(2015)Jones, Yang, Pennycook, Marshall, Van~Aert,
  Browning, Castell, and Nellist]{Jones2015-ca}
Lewys Jones, Hao Yang, Timothy~J Pennycook, Matthew S~J Marshall, Sandra
  Van~Aert, Nigel~D Browning, Martin~R Castell, and Peter~D Nellist.
\newblock Smart align—a new tool for robust non-rigid registration of
  scanning microscope data.
\newblock \emph{Adv. Struct. Chem. Imaging}, 1\penalty0 (1):\penalty0 8,
  10~December 2015.

\bibitem[Sang and LeBeau(2014)]{Sang2014-wn}
Xiahan Sang and James~M LeBeau.
\newblock Revolving scanning transmission electron microscopy: correcting
  sample drift distortion without prior knowledge.
\newblock \emph{Ultramicroscopy}, 138:\penalty0 28--35, 1~March 2014.

\bibitem[Alampara et~al.(2025)Alampara, Schilling-Wilhelmi, Ríos-García,
  Mandal, Khetarpal, Grover, Krishnan, and Jablonka]{Alampara2025-bd}
Nawaf Alampara, Mara Schilling-Wilhelmi, Martiño Ríos-García, Indrajeet
  Mandal, Pranav Khetarpal, Hargun~Singh Grover, N~M~Anoop Krishnan, and
  Kevin~Maik Jablonka.
\newblock Probing the limitations of multimodal language models for chemistry
  and materials research.
\newblock \emph{Nat. Comput. Sci.}, 5\penalty0 (10):\penalty0 952--961,
  11~October 2025.

\bibitem[Li et~al.(2026)Li, Miao, Krishnan, and Padman]{Li2026-ve}
Yubo Li, Yidi Miao, Ramayya Krishnan, and Rema Padman.
\newblock {JEV}-as-a-judge: Accept when confident, escalate when unsure.
\newblock \emph{arXiv [cs.AI]}, 22~September 2026.

\bibitem[Khan et~al.(2023)Khan, Lee, Huang, and Clark]{Khan2023-eh}
Abid Khan, Chia-Hao Lee, Pinshane~Y Huang, and Bryan~K Clark.
\newblock Leveraging generative adversarial networks to create realistic
  scanning transmission electron microscopy images.
\newblock \emph{Npj Comput. Mater.}, 9\penalty0 (1):\penalty0 85, 29~May 2023.

\bibitem[Madsen and Susi(2021)]{Madsen2021-bu}
Jacob Madsen and Toma Susi.
\newblock The {abTEM} code: transmission electron microscopy from first
  principles.
\newblock \emph{Open Res. Eur.}, 1\penalty0 (24):\penalty0 24, 24~March 2021.

\bibitem[Hjorth~Larsen et~al.(2017)Hjorth~Larsen, Jørgen~Mortensen, Blomqvist,
  Castelli, Christensen, Dułak, Friis, Groves, Hammer, Hargus, Hermes,
  Jennings, Bjerre~Jensen, Kermode, Kitchin, Leonhard~Kolsbjerg, Kubal,
  Kaasbjerg, Lysgaard, Bergmann~Maronsson, Maxson, Olsen, Pastewka, Peterson,
  Rostgaard, Schiøtz, Schütt, Strange, Thygesen, Vegge, Vilhelmsen, Walter,
  Zeng, and Jacobsen]{Hjorth-Larsen2017-us}
Ask Hjorth~Larsen, Jens Jørgen~Mortensen, Jakob Blomqvist, Ivano~E Castelli,
  Rune Christensen, Marcin Dułak, Jesper Friis, Michael~N Groves, Bjørk
  Hammer, Cory Hargus, Eric~D Hermes, Paul~C Jennings, Peter Bjerre~Jensen,
  James Kermode, John~R Kitchin, Esben Leonhard~Kolsbjerg, Joseph Kubal,
  Kristen Kaasbjerg, Steen Lysgaard, Jón Bergmann~Maronsson, Tristan Maxson,
  Thomas Olsen, Lars Pastewka, Andrew Peterson, Carsten Rostgaard, Jakob
  Schiøtz, Ole Schütt, Mikkel Strange, Kristian~S Thygesen, Tejs Vegge, Lasse
  Vilhelmsen, Michael Walter, Zhenhua Zeng, and Karsten~W Jacobsen.
\newblock The atomic simulation environment-a python library for working with
  atoms.
\newblock \emph{J. Phys. Condens. Matter}, 29\penalty0 (27):\penalty0 273002,
  12~July 2017.

\bibitem[Lobato and Van~Dyck(2014)]{Lobato2014-uu}
I~Lobato and D~Van~Dyck.
\newblock An accurate parameterization for scattering factors, electron
  densities and electrostatic potentials for neutral atoms that obey all
  physical constraints.
\newblock \emph{Acta Crystallogr. A Found. Adv.}, 70\penalty0 (6):\penalty0
  636--649, 1~November 2014.

\bibitem[Kwon et~al.(2023)Kwon, Li, Zhuang, Sheng, Zheng, Yu, Gonzalez, Zhang,
  and Stoica]{Kwon2023-gg}
Woosuk Kwon, Zhuohan Li, Siyuan Zhuang, Ying Sheng, Lianmin Zheng, Cody~Hao Yu,
  Joseph Gonzalez, Hao Zhang, and Ion Stoica.
\newblock Efficient memory management for large language model serving with
  {PagedAttention}.
\newblock In \emph{Proceedings of the 29th Symposium on Operating Systems
  Principles}, pages 611--626, New York, NY, USA, 23~October 2023. ACM.

\bibitem[Wilson(1927)]{Wilson1927-dc}
Edwin~B Wilson.
\newblock Probable inference, the law of succession, and statistical inference.
\newblock \emph{J. Am. Stat. Assoc.}, 22\penalty0 (158):\penalty0 209--212,
  June 1927.

\end{thebibliography}

\ifarxiv
  \clearpage
  \begin{center}
    {\LARGE Supplementary Information\par}
  \end{center}
  \vspace{1em}
  
\setcounter{figure}{0}
\setcounter{table}{0}
\renewcommand{\thefigure}{S\arabic{figure}}
\renewcommand{\thetable}{S\arabic{table}}
\renewcommand{\theHfigure}{S\arabic{figure}}
\renewcommand{\theHtable}{S\arabic{table}}
\renewcommand{\figurename}{Supplementary Figure}
\renewcommand{\tablename}{Supplementary Table}
\newcounter{snote}
\newcommand{\snote}[1]{\refstepcounter{snote}\section*{Supplementary Note~\thesnote: #1}}

\lstset{basicstyle=\ttfamily\small, breaklines=true, breakatwhitespace=true, breakindent=0pt, columns=fullflexible, frame=single, keepspaces=true, upquote=true}

\snote{Prompts}
\label{supp:note-prompts}
The same base prompt was used for all models and decision methods; the additions for each decision method are described below.
The system message specified the answer format and the state, that is, the type of image and the definition of the decision region.
The user message provided the target description, the image of the current FOV with the green square drawn at its center, and the question with the candidate labels, in this order.
In the listing below, the bracketed line indicates the position of the image and was not given to the model as text.
No history of previous decisions was provided.
The placeholder \texttt{\{TARGET DESCRIPTION\}} was replaced with one of the target descriptions in Supplementary Table~\ref{supp:tab-descriptions}; under the No description condition, the entire ``Target:'' paragraph was removed.

\subsection*{System message}
\begin{lstlisting}
Answer a fixed set of questions about the state the user provides. Each question lists its allowed answers; reply with exactly one label per question.

State:

You are viewing a HAADF-STEM image of the current field of view.
A line or boundary counts as inside if it passes through the green square.
\end{lstlisting}

\subsection*{User message}
\begin{lstlisting}
Target:

{TARGET DESCRIPTION}

[image: the current field of view with the green square at its center]

Question locate: Where is the target relative to the green square?

Labels:

left: the target lies outside the square, to its left
right: the target lies outside the square, to its right
up: the target lies outside the square, above it
down: the target lies outside the square, below it
inside: the target is inside the green square
none: the target is not visible in the image
\end{lstlisting}

\subsection*{Differences among the decision methods}
\begin{itemize}
  \item \TD{}: The assistant message was started with ``\texttt{locate:}'', and the next-token probabilities were read; the label with the highest probability among the six labels was taken as the answer.
  \item \SO{}: Reasoning was switched off, and the answer was constrained by a JSON schema to JSON of the form \texttt{\{"locate": "<label>"\}}.
  \item \RSO{}: Reasoning was switched on, and the model generated the same JSON answer as in \SO{} after its reasoning trace.
\end{itemize}

\subsection*{Reasoning and generation settings}
For Gemma 4, Qwen3.8, and Nemotron 3 Nano Omni, reasoning was switched by the \texttt{enable\_\allowbreak thinking} option of the chat template, which was sent to vLLM as \texttt{chat\_template\_kwargs}: it was set to \texttt{true} for \RSO{} and \texttt{false} for \SO{}.
For Qwen3.8, the reasoning effort was left at the template default (\texttt{xhigh}).
The chat template of Cosmos-Reason2, which is based on Qwen3-VL, has no such option.
Following the model card, reasoning was therefore enabled in \RSO{} by appending the following instruction to the end of the system message:
\begin{lstlisting}
Answer the question using the following format:

<think>
Your reasoning.
</think>

Write your final answer immediately after the </think> tag.
\end{lstlisting}
Without this instruction, as in \SO{}, the model answered without reasoning.

For \SO{} and \RSO{}, the vLLM server was launched with a model-specific reasoning parser (\texttt{qwen3}, \texttt{gemma4}, or \texttt{nemotron\_v3}), which separates the reasoning trace from the answer.
The JSON schema was therefore applied only to the answer after the reasoning trace, and the reasoning trace itself was not constrained.
The maximum number of output tokens, including the reasoning trace, was 8192.
The sampling parameters were taken from the \texttt{generation\_config.json} file of each model (for Cosmos-Reason2, a temperature of 0.7, top-$p$ of 0.8, and top-$k$ of 20) and were the same for \SO{} and \RSO{}.

\snote{Reasoning trace in the false-detection example}
\label{supp:note-reasoning}
This note gives the full output generated by Cosmos-Reason2 32B with \RSO{} at the first decision of the grain boundary example in Fig.~3b of the main text.
At this decision, the grain boundary was located to the right of the green square, but the model answered \texttt{inside}, and navigation was terminated as false detection.
The reasoning trace consisted of 191 tokens, and the decision took 9.8~s, shorter than the mean decision latency of \RSO{} with this model in the grain boundary task (104~s; Fig.~2b of the main text).
Although a single image was given to the model, the trace refers to it as a video.
The output is reproduced verbatim below.

\noindent\begin{minipage}{\linewidth}
\begin{lstlisting}
<think>
Okay, let's see. The user is asking where the target (a straight grain boundary in monolayer MoS2) is located relative to the green square in the image.

First, I need to recall the video. The image is a HAADF-STEM scan with a grid of white dots on a black background. The green square is centered, and the target is a boundary between two hexagonal lattices rotated by 22 degrees. The key detail here is that the grain boundary is a line or boundary that passes through the green square.

The question's labels include "inside" as an option. Since the boundary in the video is passing through the green square, that would mean part of it is inside the square. The other options like left, right, up, down, or none don't fit because the boundary isn't just adjacent but actually goes through the square. So the correct answer should be "inside"
</think>

{"locate": "inside"}
\end{lstlisting}
\end{minipage}

\section*{Supplementary Figures}

\begin{figure}[!htbp]
  \centering
  \includegraphics[width=\linewidth]{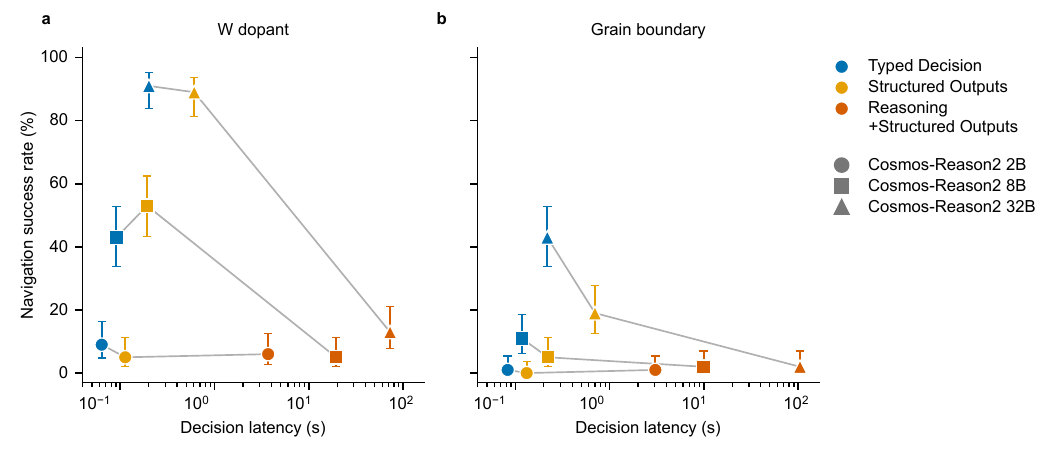}
  \caption{
      \textbf{Navigation success rate and decision latency of Cosmos-Reason2 with different model sizes.}
      The 2B, 8B, and 32B models were compared using \TD{}, \SO{}, and \RSO{} for W dopant and grain boundary navigation.
      The evaluation conditions were the same as in Fig.~2 of the main text: each condition consisted of 100 episodes with randomized starting positions and was evaluated without noise and with a step limit of 60 steps.
      Error bars indicate 95\% Wilson confidence intervals, and gray lines connect the three decision methods for the same model size.
      \textbf{a}, W dopant navigation. \textbf{b}, Grain boundary navigation.
  }
  \label{supp:fig-size}
\end{figure}

\clearpage
\section*{Supplementary Tables}

\begin{table}[htbp]
  \centering
  \caption{\textbf{VLM checkpoints used in this study.}
  All weights are publicly available on Hugging Face and were used without task-specific training.
  NVFP4 denotes versions quantized by NVIDIA in its 4-bit floating-point format.
  Cosmos-Reason2 2B and 8B were used only for the model-size comparison (Supplementary Fig.~\ref{supp:fig-size}).}
  \label{supp:tab-models}
  \small
  \begin{tabularx}{\linewidth}{>{\raggedright\arraybackslash}p{0.25\linewidth}>{\raggedright\arraybackslash\footnotesize}X>{\raggedright\arraybackslash}p{0.09\linewidth}}
    \toprule
    Model & Hugging Face repository & Precision \\
    \midrule
    Cosmos-Reason2 32B & \url{https://huggingface.co/nvidia/Cosmos-Reason2-32B} & BF16 \\
    Qwen3.8 27B & \url{https://huggingface.co/nvidia/Qwen3.8-27B-NVFP4} & NVFP4 \\
    Gemma 4 & \url{https://huggingface.co/nvidia/Gemma-4-26B-A4B-NVFP4} & NVFP4 \\
    Nemotron 3 Nano Omni & \url{https://huggingface.co/nvidia/Nemotron-3-Nano-Omni-30B-A3B-Reasoning-NVFP4} & NVFP4 \\
    \midrule
    Cosmos-Reason2 2B & \url{https://huggingface.co/nvidia/Cosmos-Reason2-2B} & BF16 \\
    Cosmos-Reason2 8B & \url{https://huggingface.co/nvidia/Cosmos-Reason2-8B} & BF16 \\
    \bottomrule
  \end{tabularx}
\end{table}

\begin{table}[htbp]
  \centering
  \caption{\textbf{Target descriptions.}
  Each description followed ``Target:'' in the prompt (Supplementary Note~\ref{supp:note-prompts}) and was compared in Fig.~4c of the main text.
  Reference was used in all experiments except those in Fig.~4c.
  The wording is shown exactly as given to the models.}
  \label{supp:tab-descriptions}
  \small
  \begin{tabularx}{\linewidth}{>{\raggedright\arraybackslash}p{0.16\linewidth}>{\raggedright\arraybackslash}X>{\raggedright\arraybackslash}X}
    \toprule
    Condition & W dopant & Grain boundary \\
    \midrule
    Reference & A single heavy dopant atom (W substituting Mo in monolayer \ce{MoS2}) that appears much brighter than all the other atomic columns. & A straight grain boundary in monolayer \ce{MoS2} where two hexagonal atomic lattices meet, with their atomic rows rotated by about 22 degrees relative to each other. \\
    Paraphrased & One tungsten atom that has replaced a molybdenum atom in single-layer \ce{MoS2}; its atomic column is far brighter than every other column in the image. & A linear interface between two grains of single-layer \ce{MoS2}, whose hexagonal atomic lattices differ in orientation by roughly 22 degrees. \\
    Name only & A W dopant in monolayer \ce{MoS2}. & A grain boundary in monolayer \ce{MoS2}. \\
    Visual only & A single atomic column in monolayer \ce{MoS2} that is much brighter than all the others. & A straight line separating two regions of a hexagonal atomic lattice in monolayer \ce{MoS2}, with atomic rows differing in direction by about 22 degrees. \\
    Technical only & A substitutional W dopant on a Mo site (W\_Mo) in monolayer \ce{MoS2}. & A $\Sigma$7 (\textasciitilde21.8$^\circ$) tilt grain boundary in monolayer \ce{MoS2}. \\
    No description & (no ``Target:'' paragraph) & (no ``Target:'' paragraph) \\
    \bottomrule
  \end{tabularx}
\end{table}

\fi

\end{document}